\documentclass[12pt]{article}

\usepackage{graphicx}
\begin{document}

\begin{center}
{\bf Dyonic black holes with nonlinear logarithmic electrodynamics } \\
\vspace{5mm} S. I. Kruglov
\footnote{E-mail: serguei.krouglov@utoronto.ca}
\underline{}
\vspace{3mm}

\textit{ Department of Physics, University of Toronto, \\60 St. Georges St.,
Toronto, ON M5S 1A7, Canada\\
Department of Chemical and Physical Sciences,\\ University of Toronto Mississauga,\\
3359 Mississauga Rd. N., Mississauga, ON L5L 1C6, Canada}
\vspace{5mm}
\end{center}

\begin{abstract}
 A new dyonic solution for black holes with spherically symmetric configurations in general relativity is  obtained. We study black holes possessing electric and magnetic charges, and the source of the gravitational field is electromagnetic fields obeying the logarithmic electrodynamics. This particular form of nonlinear electrodynamics is of interest because of its simplicity. Corrections to Coulomb's law and Reissner$-$Nordstr\"{o}m solution are found. We calculate the Hawking temperature of black holes and show that  second-order phase transitions occur for some parameters of the model.
\end{abstract}

\section{Introduction}

Dyonic black hole (BH) solutions were obtained in string theories \cite{Mignemi0}-\cite{Mignemi1} and supergravity \cite{Chamseddine}-\cite{Meessen}, gravity's rainbow \cite{Panahiyan} and dilatonic gravity \cite{Poletti}, \cite{Shapere}. The thermodynamics of BH was investigated in \cite{Pang}, \cite{Bravo}
Dyonic BHs are of interest because they have applications in different areas. We mention the Hall conductivity and the Nernst effect within  AdS/CFT correspondence \cite{Hartnoll}, \cite{Hartnoll1}, relativistic magnetohydrodynamics \cite{Caldarelli}, superconductivity \cite{Albash}, \cite{Goldstein}, and phase transitions \cite{Dutta}, \cite{Hendi}.
The nonlinear electrodynamics (NED) can solve the problems of singularities of charges at the center of particles and the infinite self-energy at the classical level. The first successful model of NED
which can solve singularity problems was proposed by Born and Infeld (BI) \cite{Born}. Then some models were proposed that have similar properties \cite{Soleng}-\cite{Kruglov1}. Heisenberg and Euler shown that NED appears from QED due to quantum corrections \cite{Heisenberg}. In \cite{Pellicer}-\cite{Quiros}, NEDs coupled to general relativity (GR) was considered. Corrections to the Reisner$-$Nordstr\"{o}m (RN) solutions and thermodynamics of BHs were studied \cite{Hendi1}-\cite{Kruglov2}. Both magnetic and electric charged BHs were studied \cite{Yajima}-\cite{Kruglov3}. It was shown that phase transitions can occur in BH.
In addition, NED coupled to GR can explain inflation and current acceleration of the universe \cite{Garcia}-\cite{Kruglov4}. In this paper we study dyonic BH (with magnetic and electric charges) in the framework of logarithmic electrodynamics. It should be noted that for weak fields (to the second order) both BI \cite{Born} and the Euler$-$Heisenberg \cite{Heisenberg} actions can be represented by logarithmic electrodynamics. Therefore, this particular NED is of interest due to its simplicity. One can consider this NED as a toy-model which allows us to study dyonic BH solutions, solutions with naked singularities,
extremal BH solutions, and thermodynamics.

The paper is organised as follows. The model of logarithmic electrodynamics is considered in Section 2. In Section 3 we obtain the dyonic solution for a BH in GR where the source is logarithmic electrodynamics. Corrections to Coulomb's law and RN solutions are found. BH thermodynamics is considered in Section 4. The Hawking temperature of a BH and heat capacity are calculated. We show that at some model parameters second-order phase transitions occur. Section 5 is a conclusion.

We use units with $c=k_B=1$, and the signature $\mbox{diag}(-,+,+,+)$.

\section{The model of logarithmic electrodynamics}

We start with the Lagrangian density of NED \cite{Soleng}
\begin{equation}
{\cal L} = -\beta^2\ln\left(1+\frac{{\cal F}}{\beta^2}\right),
 \label{1}
\end{equation}
where ${\cal F}=(1/4)F_{\mu\nu}F^{\mu\nu}=(\textbf{B}^2-\textbf{E}^2)/2$ and $F_{\mu\nu}=\partial_\mu A_\nu-\partial_\nu A_\mu$ is the field strength tensor. The parameter $\beta$ has the dimension (length)$^{-2}$ and ${\cal F}/\beta^2$ is dimensionless. As ${\cal F}\rightarrow 0$ the Lagrangian density (1) converts to the
Maxwell Lagrangian density ($-{\cal F}$).
Here, we discuss logarithmic electrodynamics in the framework of special relativity.
Using (1), we obtain the equations of motion
\begin{equation}
\partial_\mu\left(\frac{F^{\mu\nu}}{1+{\cal F}/\beta^2}\right)=0.
\label{2}
\end{equation}
From Eq. (2) with the help of the electric displacement field
$\textbf{D}=\partial{\cal L}/\partial \textbf{E}$ we find
\begin{equation}
\textbf{D}=\varepsilon\textbf{E},~~~~\varepsilon=\frac{1}{{1+\cal F}/\beta^2}.
\label{3}
\end{equation}
The magnetic field can be found from the relation $\textbf{H}=-\partial{\cal L}/\partial \textbf{B}$,
\begin{equation}
\textbf{B}= \mu\textbf{H},~~~~\mu=\varepsilon^{-1},
\label{4}
\end{equation}
and $\varepsilon\mu=1$. As a result, the speed of electromagnetic waves in vacuum is equal to the speed of light.
Equation (2) can be written as the first pair of Maxwell's equations
\begin{equation}
\nabla\cdot \textbf{D}= 0,~~~~ \frac{\partial\textbf{D}}{\partial
t}-\nabla\times\textbf{H}=0.
\label{5}
\end{equation}
The second pair of Maxwell's equation follows from the Bianchi identity $\partial_\mu \tilde{F}^{\mu\nu}=0$ ($\tilde{F}^{\mu\nu}$ is a dual tensor),
\begin{equation}
\nabla\cdot \textbf{B}= 0,~~~~ \frac{\partial\textbf{B}}{\partial
t}+\nabla\times\textbf{E}=0.
\label{6}
\end{equation}
The NED equations (2) with the parameter $\beta$ are represented in the form of nonlinear Maxwell's equations (5) and (6). The electric permittivity $\varepsilon$ and the magnetic permeability $\mu=\varepsilon^{-1}$ depend on the fields $\textbf{E}$ and $\textbf{B}$.
From Eqs. (3) and (4) we obtain the relation $\textbf{D}\cdot\textbf{H}=\varepsilon^2\textbf{E}\cdot\textbf{B}$, and as a result \cite{Gibbons}, the dual symmetry is broken because $\textbf{D}\cdot\textbf{H}\neq\textbf{E}\cdot\textbf{B}$.
In Maxwel's electrodynamics and BI electrodynamics the dual symmetry between electric and magnetic fields holds. But in QED due to loop corrections the dual symmetry is violated.
The symmetrical energy-momentum tensor, obtained from Eq. (1), is given by
\begin{equation}
T^{\mu}_{\nu}={\cal L}_{\cal F}F^{\mu}_{\alpha}F_{\nu}^{\alpha}-\delta^{\mu}_{\nu}{\cal L},
\label{7}
\end{equation}
where ${\cal L}_{\cal F}=\partial{\cal L}/\partial{\cal F}$.

In Gaussian units, at $\textbf{B}=\textbf{H}=0$, when the source is the pointlike charge, we obtain
\begin{equation}
\nabla\cdot \textbf{D}=4\pi q_e\delta(\textbf{r}).
\label{8}
\end{equation}
 The solution to Eq. (8) is given by
\begin{equation}
\textbf{D}=\frac{q_e}{r^3}\textbf{r}.
\label{9}
\end{equation}
Making use of Eq. (3) we obtain from Eq. (9)
\begin{equation}
E\left(\frac{1}{1-E^2/(2\beta^2)}\right)=\frac{q_e}{r^2}.
\label{10}
\end{equation}
At the center, $r\rightarrow 0$, Eq.(10) has the solution
\begin{equation}
E_{max}=\sqrt{2}\beta.
\label{11}
\end{equation}
Thus, a maximum of the electric field at the center of charges is finite and equals $E_{max}=\sqrt{2}\beta$. A similar property occurs in BI electrodynamics.

\section{Nonlinear electrodynamics and the dyonic solution}

Let us consider logarithmic electrodynamics coupled to GR with the action
\begin{equation}
I=\int d^4x\sqrt{-g}\left(\frac{1}{2\kappa^2}R+ {\cal L}\right),
\label{12}
\end{equation}
where $\kappa^2=8\pi G\equiv M_{Pl}^{-2}$, $G$ is Newton's constant, $M_{Pl}$ is the reduced Planck mass, and ${\cal L}$ is given by Eq. (1). We suppose that the metric is static, spherically
symmetric and is given by
\begin{equation}
ds^2=-A(r)dt^2+\frac{1}{A(r)}dr^2+r^2(d\vartheta^2+\sin^2\vartheta d\phi^2).
\label{13}
\end{equation}
This metric is realised using the Einstein equations, if the stress tensor obeys the equality
$T^0_0=T^r_r$. Then the field equations lead to \cite{Bronnikov3}, \cite{Bronnikov4}\footnote{We use a little different notations compared to \cite{Bronnikov3}, \cite{Bronnikov4}.}
\begin{equation}
B=\frac{q_m}{r^2},~~~~E^2=\frac{q_e^2}{{\cal L}^2_{\cal F}r^4}=\frac{q^2_e}{r^4}\left(1+\frac{{\cal F}}{\beta^2}\right)^2,
\label{14}
\end{equation}
\begin{equation}
{\cal F}=\frac{q_m^2}{2r^4}-\frac{q_e^2}{2r^4}\left(1+\frac{{\cal F}}{\beta^2}\right)^2,
\end{equation}
\label{15}
where $q_e$ and $q_m$ are electric and magnetic charges, respectively. The dyonic solution to the quadratic equation (15) for $E^2$ is given by
\begin{equation}
E^2=2\beta^2\left(1+\frac{a}{r^2}+\frac{r^4}{2b}-\frac{\sqrt{r^8+4b(a+r^4)}}{2b}\right),
\end{equation}
\label{16}
\begin{equation}
a=\frac{q_m^2}{2\beta^2},~~~~b=\frac{q_e^2}{2\beta^2}.
\label{17}
\end{equation}
It follows from Eqs. (16) and (17) that if $q_m\neq 0$ the electric field at the origin ($r=0$) possesses a singularity. However, when the magnetic charge is zero ($q_m=0$), the singularity vanishes. In this case, making use of Taylor series, we obtain from Eqs. (16) and (17) (as $r\rightarrow 0$)
\begin{equation}
E=\sqrt{2}\beta-\frac{\beta^2r^2}{q_e}+\frac{\beta^3r^4}{2\sqrt{2}q_e^2}+{\cal O}(r^6)~~~~(q_m=0).
\label{18}
\end{equation}
From (18) one finds the result (11) that the maximum of the electric field at the center ($r=0$) is $E_{max}=\sqrt{2}\beta$.
If both the electric and magnetic charges are nonzero, we obtain from (16) the series as $r\rightarrow\infty$
\begin{equation}
E=\frac{q_e}{r^2}+\frac{q_e(q_m^2-q_e^2)}{2\beta^2 r^6}-\frac{q_e^3(q_m^2-q_e^2)}{2\beta^4r^{10}}+{\cal O}(r^{-11}).
\label{19}
\end{equation}
Equation (19) contains corrections to Coulomb's law. It is interesting that if we consider the self-dual solution, $q_m=q_e$, corrections to Coulomb's law vanish.

From Eq. (7) we obtain the energy density
\[
\rho(r)\equiv T^0_0={\cal L}_{\cal F}F^{0}_{n}F_0^{n}-{\cal L}
\]
\begin{equation}
=\beta^2\left(\sqrt{1+\frac{4ab}{r^8}+\frac{4b}{r^4}}-1\right)+\beta^2\ln\left(\frac{\sqrt{r^8+4ab+
4br^4}-r^4}{2b}\right).
\label{20}
\end{equation}
It should be noted that the energy density (20) at the center ($r\rightarrow 0$) possesses a singularity ($\rho\rightarrow \infty$).
The metric function $A(r)$ in Eq. (13), according to the Einstein
equations, is given by \cite{Bronnikov}
\begin{equation}
A(r) = 1-\frac{2M(r)G}{r}.
\label{21}
\end{equation}
Here, the mass function is
\begin{equation}
M(r)=m-\int_r^\infty\rho(r)r^2dr,
\label{22}
\end{equation}
where $m$ is the total mass of a BH. It should be noted that the BH mass $m$ is a free parameter.

Making use of Eqs. (20) and (22), we obtain
\[
M(r)=m-\int_r^\infty \beta^2r^2\biggl(\sqrt{1+\frac{4ab}{r^8}+\frac{4b}{r^4}}-1
\]
\begin{equation}
+\ln\left(\frac{\sqrt{r^8+4ab+4br^4}-r^4}{2b}\right)\biggr)dr.
\label{23}
\end{equation}
To estimate the mass function as $r\rightarrow\infty$, we use the energy density (20) as $r\rightarrow\infty$,
\begin{equation}
\rho(r)=\beta^2\left(\frac{a+b}{r^4}-\frac{(b-a)^2}{2r^8}+\frac{(b-a)(2b^2-ab-a^2)}{3r^{12}}\right)+{\cal O}(r^{-13}).
\label{24}
\end{equation}
Using (23), we obtain the mass function as $r\rightarrow\infty$,
\begin{equation}
M(r)=m-\beta^2\left(\frac{a+b}{r}-\frac{(b-a)^2}{10r^5}+\frac{(b-a)(2b^2-ab-a^2)}{27r^{9}}\right)+{\cal O}(r^{-10}).
\label{25}
\end{equation}
As a result, Eqs. (17), (21) and (25) lead to an asymptotic form of the metric function as $r\rightarrow\infty$,
\[
A(r)=1-\frac{2mG}{r}+\frac{(q_e^2+q_m^2)G}{r^2}-\frac{(q_e^2-q_m^2)^2G}{20\beta^2r^6}
\]
\begin{equation}
+\frac{(q_e^2-q_m^2)(2q_e^4-q_e^2q_m^2-q_m^4)G}{108\beta^4r^{10}}+{\cal O}(r^{-11}).
\label{26}
\end{equation}
The first three terms in Eq. (26) give the Reissner$-$Nordstr\"{o}m solution, and other terms represent  corrections. At $q_e=q_m$, the corrections to the Reissner$-$Nordstr\"{o}m solution disappear.
Using Eqs. (21) and (23) and introducing unitless variables $z=\beta r^2/q_e$ and $n=q_m^2/q_e^2$, the metric function (21) becomes
\[
A(z)=1-\frac{q_e\beta G}{\sqrt{z}}\biggl[\frac{2m}{g_e^{3/2}\sqrt{\beta}}
\]
\begin{equation}
-\int_z^\infty\biggl(\sqrt{z+\frac{2}{z}+\frac{n}{z^3}}-
\sqrt{z}+\sqrt{z}\ln\left(\sqrt{z^4+2z^2+n}-z^2\right)\biggr)dz \biggr].
\label{27}
\end{equation}
At large $z$, the metric function (27), in terms of unitless values, is given by
\[
A(z)=1-B\biggl[\frac{C}{\sqrt{z}}-\frac{n+1}{z}+\frac{(n-1)^2}{20z^3}
\]
\begin{equation}
 -\frac{(n-1)(n^2+n-2)}{108z^5}\biggr],~~~~B\equiv q_e\beta G,~~~~C\equiv\frac{2m}{g_e^{3/2}\sqrt{\beta}}.
\label{28}
\end{equation}
The metric function $A(z)$, for different values of $B$, $C$ and $n=q_m^2/q_e^2$, is plotted in Figs. (1) and (2).
\begin{figure}[h]
\includegraphics[height=3.0in,width=3.0in]{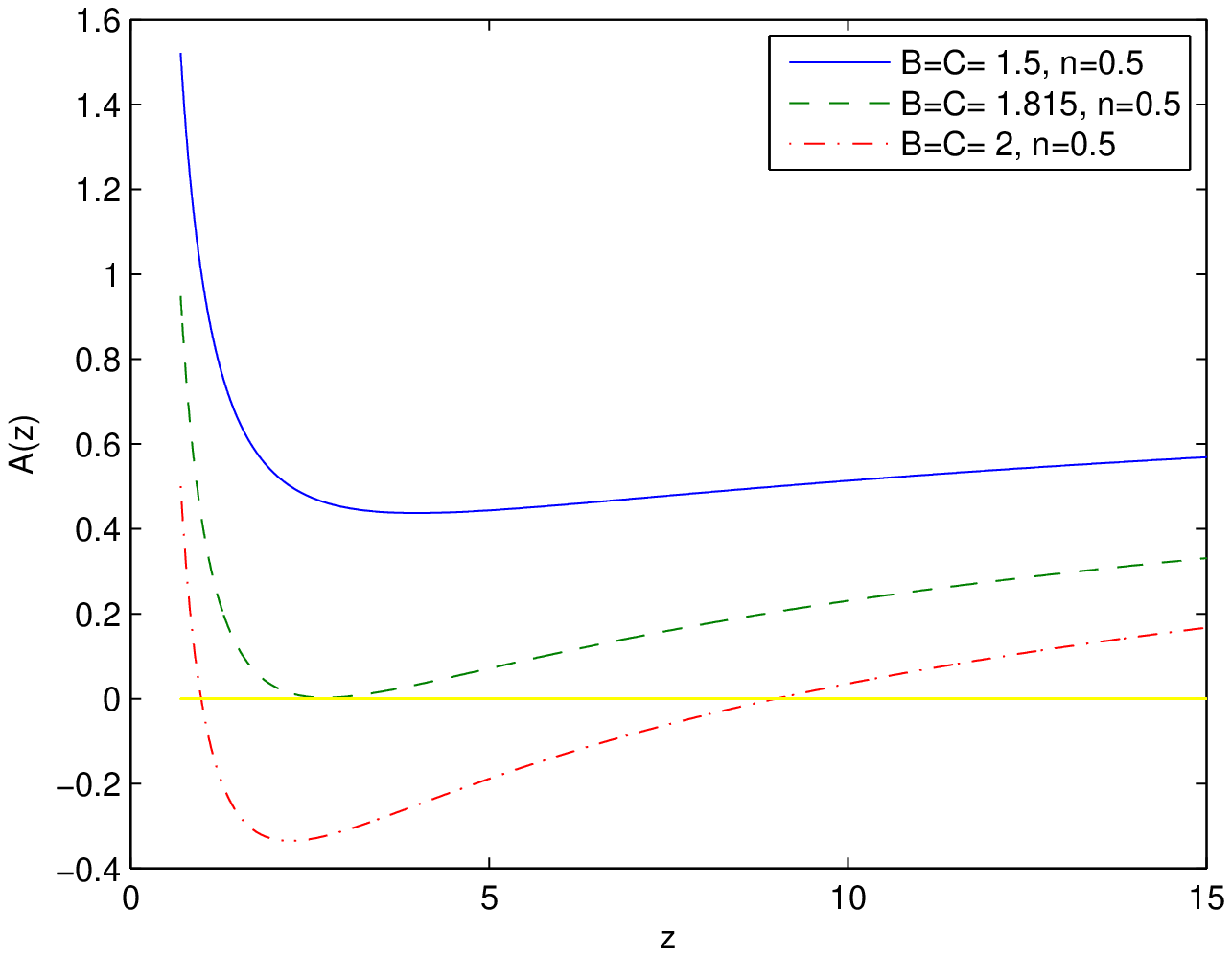}
\caption{\label{fig.1}The metric function $A(z)$ for $n=0.5$. The solid line corresponds to $B=C=1.5$, the dashed line to $B=C=1.815$, and the dashed-dotted line to $B=C=2$.}
\end{figure}
\begin{figure}[h]
\includegraphics[height=3.0in,width=3.0in]{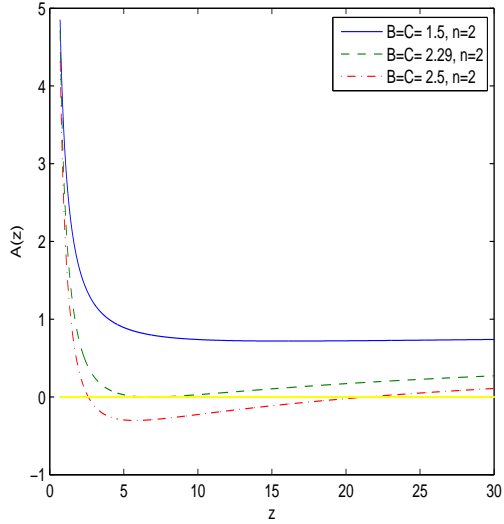}
\caption{\label{fig.2}The metric function $A(z)$ for $n=2$. The solid line corresponds to $B=C=1.5$, the dashed line to $B=C=2.29$, and the dashed-dotted line to $B=C=2.5$.}
\end{figure}

The figures show that at some parameters there can be BH solutions with horizons, solutions with naked singularities, and extremal BH solutions. Thus, according to Fig. 1, at $B=C=1.5$ ($q_e^2=2q^2_m$) one has a naked singularity, at $B=C=1.815$ ($q_e^2=2q_m^2$) we have an extremal BH solution, and at $B=C=2$ ($q_e^2=2q^2_m$) there is a BH solution with two horizons. Fig. 2 shows the similar behaviour of $A(z)$ for $q_e^2=0.5q^2_m$. As a result, for $q_m>q_e$ the event horizon radius is larger than in the case $q_e>q_m$.

\section{Thermodynamics}

The Hawking temperature of a BH is given by
\begin{equation}
T_H=\frac{\kappa_S}{2\pi}=\frac{A'(r_+)}{4\pi},
\label{29}
\end{equation}
were $\kappa_S$ is the surface gravity and $r_+$ is the horizon radius. Thus we assume the existence of the event horizon for some parameters $\beta$, $q_e$, $q_m$, and $m$.  From Eqs. (21) and (22) we obtain the relations 
\begin{equation}
A'(r)=\frac{2 GM(r)}{r^2}-\frac{2GM'(r)}{r},~~~M'(r)=r^2\rho(r),~~~M(r_+)=\frac{r_+}{2G}.
\label{30}
\end{equation}
Making use of Eqs. (29) and (30), one finds
\begin{equation}
T_H=\frac{1}{4\pi}\left(\frac{1}{r_+}-2Gr_+\rho(r_+)\right).
\label{31}
\end{equation}
With the help of Eqs. (17), (20)  and (31), and introducing the unitless variables $z_+=\beta r_+^2/q_e$, $n=q_m^2/q_e^2$, we obtain the Hawking temperature
\[
T_H=\frac{\sqrt{\beta}}{4\pi\sqrt{q_e}}\biggl[\frac{1}{\sqrt{z_+}}-2G\beta q_e\sqrt{z_+}\biggl(\sqrt{1+\frac{n}{z_+^4}+\frac{2}{z_+^2}}-1
\]
\begin{equation}
+\ln\left(\sqrt{z_+^4+n+2z_+^2}-z_+^2\right)\biggr)\biggr].
\label{32}
\end{equation}
It should be noted that the event horizon radius $r_+$ depends on the parameter $\beta$. Using the relation $2GM(r_+)=r_+$, we find from Eq. (23) the dependence of the horizon radius on the parameter $\beta$:
\begin{equation}
G\beta=\frac{\sqrt{z_+}}{\frac{2m}{\sqrt{q_e\beta}}-q_e\int_{z_+}^\infty\left[\sqrt{z^2+\frac{n}{z^2}+2}-z+
z\ln\left(\sqrt{z^4+n+2z^2}-z^2\right)\right]\frac{dz}{\sqrt{z}}}.
\label{33}
\end{equation}
Substituting $G\beta$ from (33) into (32), we obtain the BH Hawking temperature as
\[
T_H=\frac{\sqrt{\beta}}{4\pi\sqrt{q_e}}\biggl(\frac{1}{\sqrt{z_+}}
\]
\begin{equation}
-\frac{2\left(\sqrt{z_+^2+\frac{n}{z_+^2}+2}-z_++z_+\ln\left(\sqrt{z_+^4+n+2z_+^2}-z_+^2\right)\right)}
{\frac{2m}{q_e^{3/2}\sqrt{\beta}}-\int_{z_+}^\infty\left[\sqrt{z+\frac{n}{z^3}+\frac{2}{z}}-\sqrt{z}+
\sqrt{z}\ln\left(\sqrt{z^4+n+2z^2}-z^2\right)\right]dz}\biggr).
\label{34}
\end{equation}
Note that the parameter $C\equiv m/(q_e^{3/2}\sqrt{\beta})$ is unitless. The Hawking temperature versus $z_+$ is plotted in Figs. 3 and 4 for different parameters $C$ and $n$.
\begin{figure}[h]
\includegraphics[height=3.0in,width=3.0in]{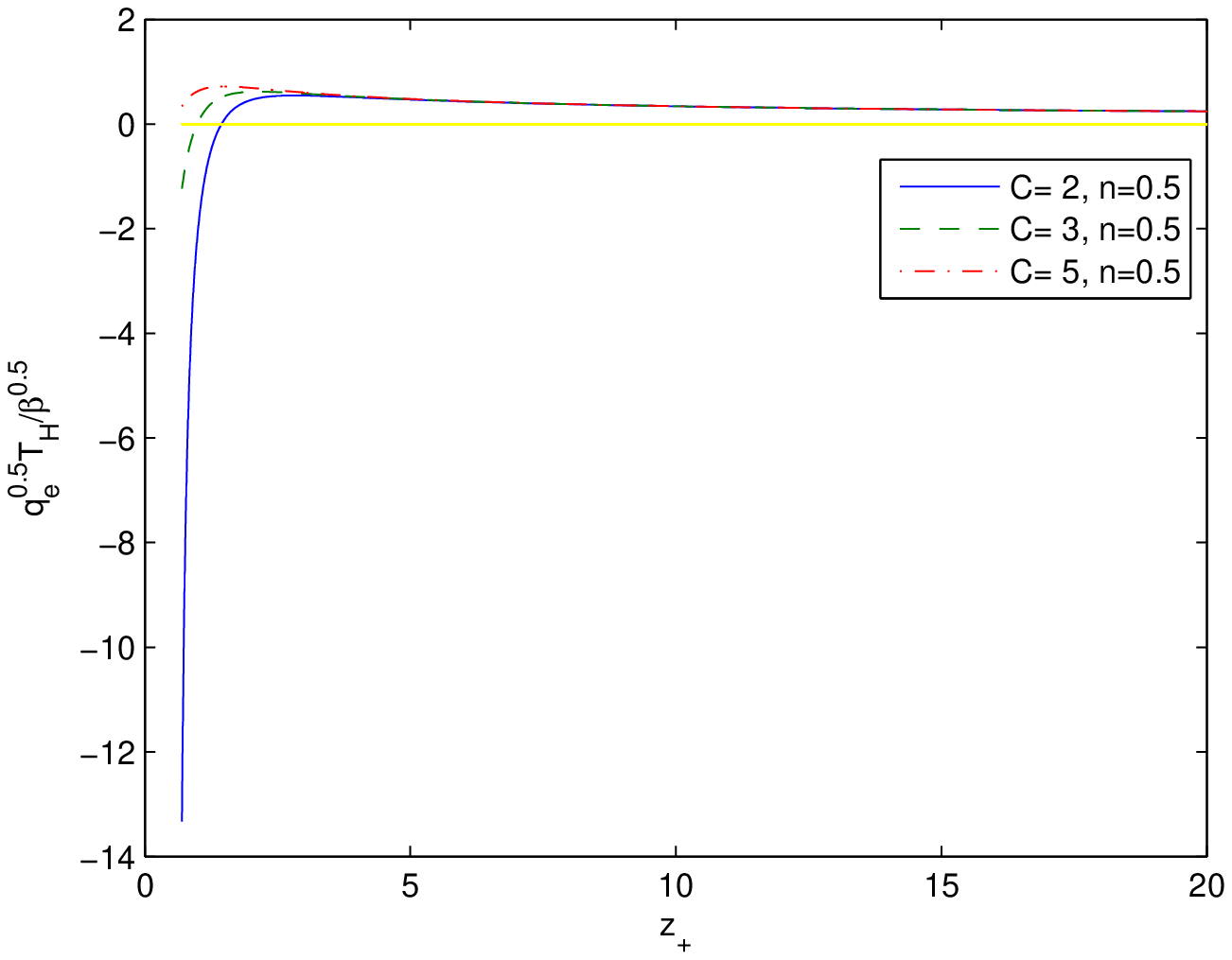}
\caption{\label{fig.3}The Hawking temperature $T_H(z_+)$ for $n=0.5$. The solid line corresponds to $C=2$, the dashed line to $C=3$, and the dashed-dotted line to $C=5$.}
\end{figure}
\begin{figure}[h]
\includegraphics[height=3.0in,width=3.0in]{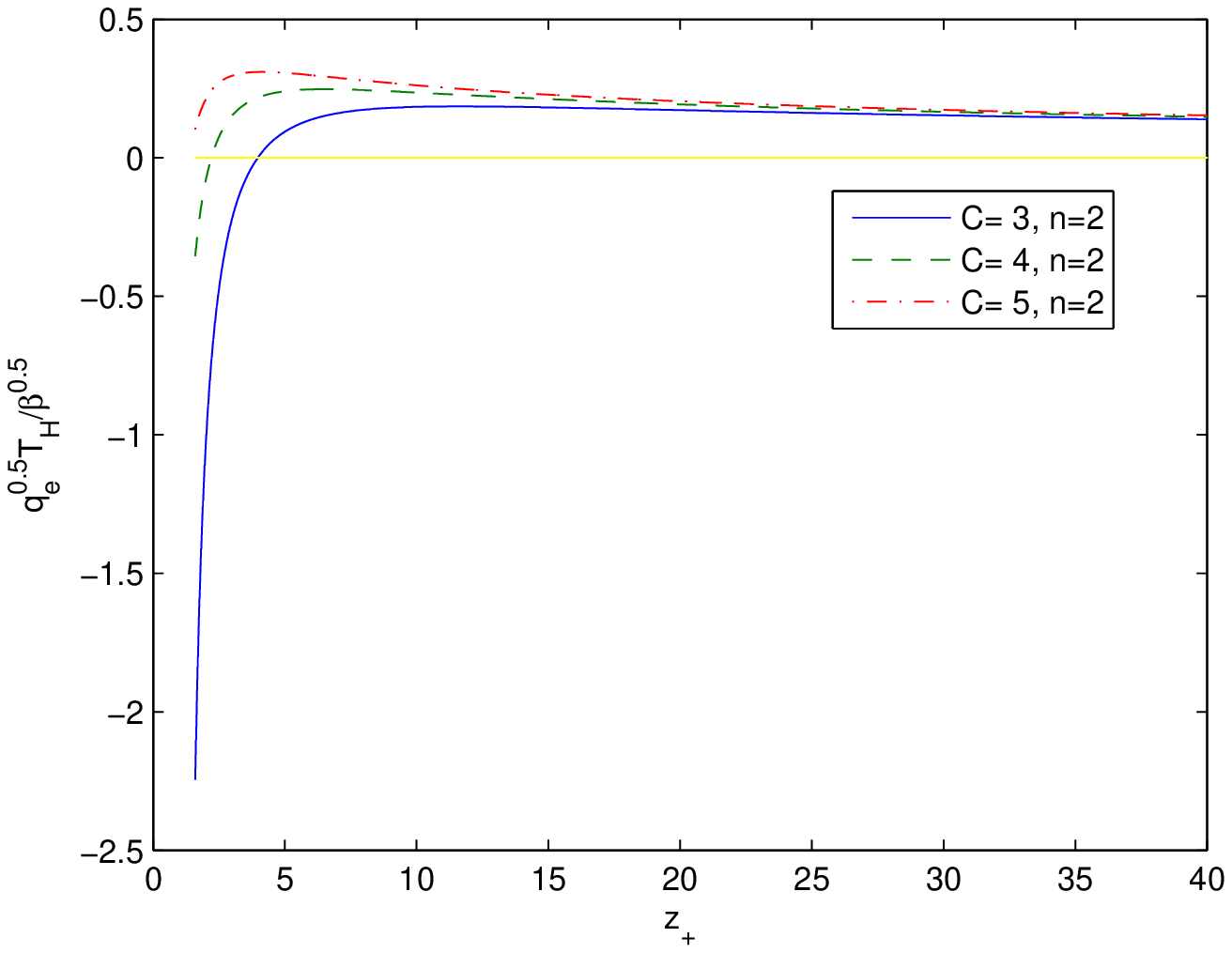}
\caption{\label{fig.4}The Hawking temperature $T_H(z_+)$ for $n=2$. The solid line corresponds to $C=3$, the dashed line to $C=4$, and the dashed-dotted line to $C=5$.}
\end{figure}

Figures show that there are maxima of Hawking temperatures at some values of horizon radius. When the temperature is negative BHs are unstable.

To investigate phase transitions, one can study the heat capacity. A second-order phase transition takes place if the heat capacity is singular. The BH entropy satisfies the Hawking area law $S=\mbox{Area}/(4G)=\pi r_+^2/G$.
Then we obtain the heat capacity
\begin{equation}
C_q=T_H\left(\frac{\partial S}{\partial T_H}\right)_q=\frac{T_H\partial S/\partial r_+}{\partial T_H/\partial r_+}=\frac{2\pi r_+T_H}{G\partial T_H/\partial r_+}.
\label{35}
\end{equation}
In accordance with (35) the heat capacity diverges if the Hawking temperature has an extremum ($\partial T_H/\partial r_+=0$), and then there is a second-order phase transition.
According to Figs. 3 and 4, we have maxima of Hawking temperatures, and therefore the second-order phase transitions occur for some values of the horizon radius $r_+$ ($z_+=\beta r_+^2/q_e$).
The expression for $C_q$ is complicated to write it down.

\section{Conclusion}

We have considered logarithmic electrodynamics (NED) with the free dimensional parameter $\beta$. At weak fields the model is converted into Maxwell's electrodynamics, i.e. the correspondence principle holds. A dual symmetry between electric and magnetic fields is broken, as in QED with loop corrections. An attractive feature of this NED is that there is no singularity of charged objects at the center of charged objects, similarly to BI electrodynamics. In addition, the total electrostatic energy of charges is finite. We have calculated corrections to Coulomb's law as $r\rightarrow\infty$. It was demonstrated that at $q_e=q_m$ corrections are absent. This NED coupled with the gravitational field was studied, and a dyonic solution for a BH in GR was obtained.
Corrections to the Reissner$-$Nordstr\"{o}m solution as $r\rightarrow\infty$, which disappear at $q_e=q_m$, were found. We have obtained expressions for the BH Hawking temperature and heat capacity. It was demonstrated  that there are second-order phase transitions for some parameters of the model.

\end{document}